\documentclass[useAMS,usenatbib]{mn2e}

 \usepackage{times}
 \usepackage{graphicx}
 \usepackage{journals}

\title[IMF determination in unresolved stellar
populations]{A new technique for the determination of the initial mass function in unresolved stellar populations}
\author[Podorvanyuk, Chilingarian \&
Katkov]{Nikolay Yu. Podorvanyuk$^{1}$\thanks{E-mail:
nicola@sai.msu.ru}, Igor V. Chilingarian$^{2,1}$, Ivan Yu. Katkov$^{1}$\\
$^{1}$Sternberg Astronomical Institute, Moscow State University, 13 Universitetski prospect, 119992 Moscow, Russia\\
$^{2}$Smithsonian Astrophysical Observatory, 
Harvard-Smithsonian Center for Astrophysics, 60 Garden St. MS09, Cambridge, MA 02138 USA}
\begin{document}

\date{Accepted 2013 March 5. Received 2013 March 5; in original form 2013 January 18}

\pagerange{\pageref{firstpage}--\pageref{lastpage}} \pubyear{2013}

\maketitle

\label{firstpage}

\begin{abstract} 
We present a new technique for the determination of the low-mass slope
($\alpha_1$; $M_* < 0.5 M_{\odot}$) of the present day stellar mass function
(PDMF) using the pixel space fitting of integrated light spectra.  It can be
used to constrain the initial mass function (IMF) of stellar systems with
relaxation timescales exceeding the Hubble time and testing the IMF
universality hypothesis.  We provide two versions of the technique: (1) a
fully unconstrained determination of the age, metallicity, and $\alpha_1$
and (2) a constrained fitting by imposing the externally determined
mass-to-light ratio of the stellar population.  We have tested our approach
by Monte-Carlo simulations using mock spectra and conclude that: (a) age,
metallicity and $\alpha_1$ can be precisely determined by applying the
unconstrained version of the code to high signal-to-noise datasets (S/N=100,
$R=7000$ yield $\Delta \alpha_1 \approx 0.1$); (b) the $M/L$ constraint
significantly improves the precision and reduces the degeneracies, however
its systematic errors will cause biased $\alpha_1$ estimates; (c) standard
Lick indices cannot constrain the PDMF because they miss most of the mass
function sensitive spectral features; (d) the $\alpha_1$
determination remains unaffected by the high-mass IMF shape ($\alpha_3$;
$M_* \ge 1 M_{\odot}$) variation for stellar systems older than 8~Gyr, while
the intermediate-mass IMF slope ($\alpha_2$; $0.5 \le M_* < 1 M_{\odot}$)
may introduce biases into the best-fitting $\alpha_1$ values if it is different 
from the canonical value $\alpha_2 = 2.3$. We analysed observed intermediate 
resolution spectra of ultracompact dwarf galaxies with our technique and 
demonstrated its applicability to real data.
\end{abstract}

\begin{keywords}
methods: data analysis -- globular clusters: general --
galaxies: star clusters: general -- galaxies: dwarf --
galaxies: kinematics and dynamics -- galaxies: stellar content
\end{keywords}

\section{Introduction}

The stellar IMF is one of the most important intrinsic properties of the star
formation process.  Its origin still remains unclear.  Historically, it was
thought to be a unimodal power law function $dN/dM \propto M^{-(\alpha-1)},
\alpha=2.35$ \citep{Salpeter55}.  The current working hypothesis
\citep{Kroupa02} of the IMF universality suggests that its shape does not
vary among different star forming regions and can be represented as a
bimodal power law with different low- and high-mass slopes ($\alpha_1 = 1.3$
for $M < 0.5\,M_{\odot}$ and $\alpha_2 = 2.3$ at larger masses).  Other
possibilities include variations of uni- or multi-modal power laws (e.g. 
\citealp{MS79}) or a lognormal function \citep{Chabrier03}.  The IMF
shape has a direct impact on galaxy evolution because it defines the number
of Wolf--Rayet stars and supernovae and therefore the amount of star
formation feedback.  The low-mass IMF slope has strong influence on stellar
mass-to-light ratios because of high $M/L$ values of dwarf late type stars.

Classical IMF determination techniques (see \citet{Kroupa02} for a review)
use direct star counts in open clusters and H{\sc ii} associations. 
However, they are affected by various phenomena such as dynamical evolution
of star clusters, variable and often strong dust extinction in star forming
regions, as well as by the low number statistics.  In this light, compact
stellar systems (CSS), that is globular clusters (GC) and ultracompact dwarf
galaxies (UCD, \citealp{Drinkwater+03}) which are orders of magnitude more
massive, free of the interstellar medium and populated by old stars present
a unique laboratory to study the IMF.

CSSs observed today might have experienced dynamical evolution effects on
their stellar mass functions, i.e.  the observed PDMF may differ
from the IMF.  It is known \citep{Spitzer87, BM03, KAS07, KM09} that in
globular clusters the dynamical evolution causes mass segregation, i.e. 
massive stars move towards the centre while low-mass stars migrate to
the cluster outskirts, where they are tidally stripped during the passages
close to the centre of a host galaxy or through its disc.  This creates a
deficit of low-mass stars in a cluster, changing the shape of its integrated
stellar mass function.  The characteristic timescale of this process is
related to the dynamical relaxation time, which can be estimated for a CSS
\citep{Mieske+08} as ${t_{\rm rel}} = \frac{0.234}{\log {M_{\rm dyn}}}
\sqrt{\frac {M_{\rm dyn}{r_{e}^3}}{0.0045}}$~Myr, where ${M_{\rm dyn}}$ is
in $M_{\odot}$ and the effective radius ${r_{e}}$ is in pc.  Sufficiently
massive GCs and UCDs have long relaxation timescales (a few Gyrs to many
$t_{\rm Hub}$) and therefore we can neglect the dynamical evolution effects
in those systems.  Hence, the PDMF there becomes a good approximation of the
IMF for low-mass stars which have not yet finished their evolution.

One has to keep in mind that the mass segregation is not the only
process which may result in the low-mass stars deficiency in CSSs.  It can
also be caused by the residual-gas expulsion from initially mass segregated
star clusters, as it was shown for galactic GCs \citep{Marks2008} and UCDs
\citep{DFK10}.  However, the dynamical analysis of Fornax cluster UCDs does
not show any signs of the lack of low-mass stars in these relatively massive
systems \citep{Mieske+08,CMHI11}.  There is also a strong relationship
between the low-mass IMF slope and the concentration in GCs \citep{DMPP}. 
It turns out to be possible to estimate the phase of the low-mass star loss
process from observations and therefore to make a conclusion about the
equivalence of PDMF and IMF for every particular CSS.

In this paper we describe a new technique of deriving the low-mass
IMF slope by using the pixel fitting of intermediate- and high-resolution
spectra of stellar systems integrated along the line of sight.  This
approach is inspired by the results of \citet{CCB08,CMHI11} where the
stellar masses derived using the \citet{Salpeter55} IMF turned to exceed the
dynamical mass estimates resulting in unphysical negative dark matter
contents.

\section{ New method: simulations and applications}

We present two versions of the technique to derive the low-mass slope of the
stellar IMF from integrated light spectra using full spectral fitting
developed as an extension of the {\sc nbursts} package \citep{CPSK07, CPSA07}.

{\sc nbursts} is an approach to determine the parametrized line-of-sight
velocity distribution (LOSVD) and star formation history (SFH) of unresolved
stellar populations by means of the full spectral fitting in the pixel
space.  Previously the {\sc nbursts} technique was mostly used with simple
stellar population (SSP) models characterized by only two parameters, age
and metallicity of the instantaneous starburst event computed for a
pre-defined IMF shape.  In this study, we added an extra dimension to the
grid of SSP models by varying the low-mass end slope $\alpha_1$ of the
Kroupa canonical IMF \citep{Kroupa02} leaving the slope above $0.5 M_{\odot}$ fixed
($\alpha_2 = 2.3$).  Hence, $\alpha_1$ becomes an additional free parameter
returned by the modified fitting procedure.  We computed stellar population
models for ``non-standard'' $\alpha_1$ values using both, high and low
resolution stellar atmosphere grids supplied with the {\sc pegase.hr}
evolutionary synthesis code \citep{LeBorgne+04}, the empirical high
resolution {\sc elodie.3.1} stellar library \citep{PSKLB07} and the low
resolution BaSeL theoretical atmospheres \citep{LCB97} used in the original
{\sc pegase.2} package \citep{FR97}.  The full spectral fitting is performed
with high-resolution models while the mass-to-light ratios of stellar
populations in broad band filters are estimated using the low-resolution
grid.  We used standard {\sc pegase.hr} settings for all other parameters:
5~per~cent of close binary systems, low and high cut-off stellar masses of
0.1 and 120 $M_{\odot}$ correspondingly.

In {\itshape the first version} of our new approach an observed spectrum is
fitted against an optimal template represented by a linear combination of
SSPs each of them characterised by the age, metallicity and the low-mass IMF
slope $\alpha_1$, determined in the same minimization loop, using the
3-dimensional cubic spline interpolation on the pre-computed grid of 
high-resolution SSP models. 

In this paper we deal with observed spectra of CSSs which can be well
represented by single component SSP models, however the code provides a
possibility of fitting multiple component stellar populations.  The model
grid has 25 nodes in age (10~Myr to 20~Gyr), 10 nodes in metallicity ([Fe/H]
from -2.5 to 1.0~dex) and 10 nodes in the low-mass IMF slope (0.4 to 3.1
with a step of 0.3).  During the fitting procedure, the models are broadened
using the Gauss-Hermite parametrization \citep{vdMF93} of the LOSVD which is
penalized toward purely Gaussian solution as proposed by \citet{CE04}.  In
this paper we consider only purely Gaussian LOSVDs.

The ${{\chi}^2}$ value (without penalization) is computed as follows:
\begin{equation}
{{\chi}^2} = 
\sum \limits_{{N_\lambda}} 
\frac{ \bigl( {F_i}-{P_{1p}}({T_i}(t,Z,\alpha_1) \otimes \mathcal{L} (v,\sigma,{h_3},{h_4})+{P_{2q}}) \bigr)^2}
{{\Delta}{{F_i}^{2}}}
\end{equation}

\noindent where ${F_i}$ and ${\Delta}{{F_i}}$ are
observed flux and its uncertainty; $\mathcal{L}$ is LOSVD used as a
convolution kernel for ${T_i}(t,Z,\alpha_1)$, the flux from a
synthetic spectrum, represented by a linear combination of SSP models
characterised by their ages $t$, metallicities $Z$, and low-mass IMF slopes
$\alpha_1$, and convolved according to the line-spread function of the
spectrograph; ${P_{1p}}$ and ${P_{2q}}$ are multiplicative and additive
Legendre polynomials of orders $p$ and $q$ for correcting the continuum;
$v$, $\sigma$, ${h_{3}}$ and ${h_{4}}$ are radial velocity, velocity
dispersion and Gauss-Hermite coefficients of the LOSVD respectively.

The ${\chi^2}$ minimization is performed using the constrained non-linear
Levenberg--Marquardt minimization implemented in the {\sc mpfit} IDL package
(by C.~Markwardt, NASA).  Since it requires the second derivatives to be
continuous, we use a three-dimensional spline interpolation of the SSP grid
to evaluate a model for a given set of ($t$, [Fe/H], $\alpha_1$).  In this
fashion, the values of $t$, [Fe/H], and $\alpha_1$ are returned by the
minimization procedure in the same loop along with the parameters of the
continua.

{\itshape The second version} of our technique is mathematically
identical to the pure {\sc nbursts} technique but the specific grid of input
SSP models is constructed and supplied for every observed spectrum.  Here we
assume that the object being studied contains no dark matter.  This
assumption is probably true for GCs and UCDs.  In this case, the dynamical
mass-to-light ratio $(M/L)_{\rm{dyn}}$ equals to the stellar mass-to-light
ratio $(M/L)_{*}$.  For a number of compact stellar systems, the
$(M/L)_{\rm{dyn}}$ values were determined from the analysis of their
internal structure and observed velocity dispersion profiles and are
available in the literature.  On the other hand, we derive the $(M/L)_{*}$
values from low-resolution stellar population models for every set of ($t$,
[Fe/H], $\alpha_1$) and for every given $t$ and [Fe/H] in the old stellar
population regime this function is monotonic.  Hence, if we know
$(M/L)_{\rm{dyn}}$ and impose the zero dark matter constraint, for every
point on the ($t$, [Fe/H]) plane we can find the value of $\alpha_1$ such as
$(M/L)_{*}$ in that point of the parameter space equals to
$(M/L)_{\rm{dyn}}$.  Thus, we can compute a grid of SSP models in the
age--metallicity space varying the low-mass IMF slope so that the
$(M/L)_{*}$ values are constant all over the grid and equal to the
``external'' $(M/L)_{\rm{dyn}}$.  Along with this grid, we will also map the
behaviour of $\alpha_1$ as a function of $t$ and [Fe/H].  Then, if we feed
this SSP grid to the standard {\sc nbursts} full spectral fitting procedure
and determine the pair of best-fitting values of age and metallicity, we
will automatically get the $\alpha_1$ value corresponding to this
best-fitting solution which will measure the low-mass PDMF slope of the
stellar population being studied.

It may happen that in order to reach the imposed value of $(M/L)$ the
low-mass IMF slope has to be outside our model grid ($0.4 < \alpha_1 < 3.1$).
In this case, we take the model corresponding to the nearest limiting
$\alpha_1$ value (maximal or minimal).

In some cases, e.g. high-resolution spectra of extragalactic GCs or UCDs
where the internal velocity dispersion can be measured from a spectrum
itself and where it uniquely defines $(M/L)_{\rm{dyn}}$, it is possible to
modify the minimization procedure and close the loop so that no external
$(M/L)$ value needs to be supplied.  However, in this case the procedure
becomes very complex and we do not consider this approach in this
paper.  As we demonstrate from simulations, velocity dispersion does
not exhibit any degeneracies with age and $\alpha_1$, therefore if $\sigma$
is determined independently, it should not affect the final $\alpha_1$
estimate.

\subsection{Full spectral fitting of mock data}

We performed Monte-Carlo simulations in order to study the precision,
stability, and possible biases of the proposed technique for the
determination of the low-mass IMF slope. We generated mock datasets (10 to
1000 noise realisations for every combination of input parameters) for the
signal-to-noise ratios of 25, 50, and 100, and the spectral resolution
values $R=4000 \dots 10000$. Then we fitted them using both versions of our
technique.

\begin{figure*}
\includegraphics[width=\hsize]{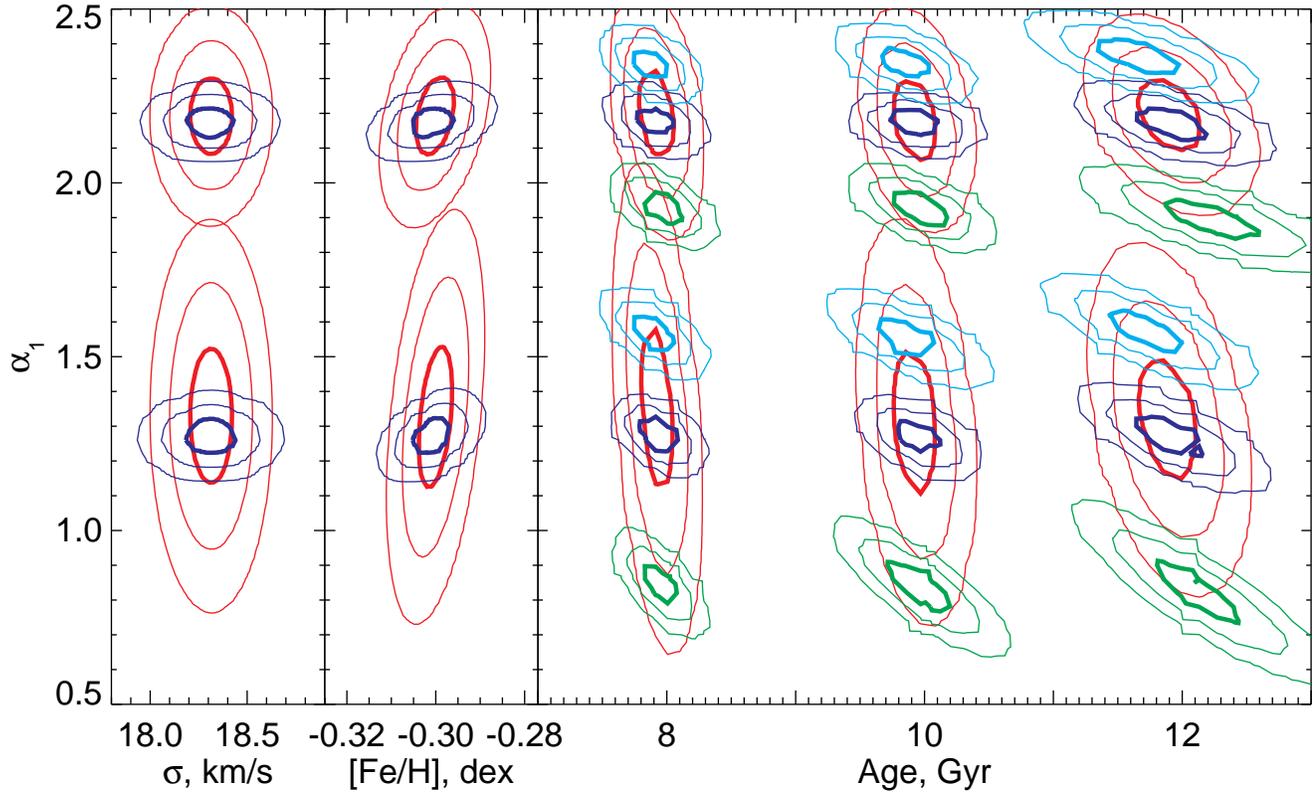}
 \caption{Results of the full spectral fitting of mock spectra. Red and blue contours correspond
to the first (all free parameters) and the second (constrained M/L) versions
of the technique. Cyan and green contours demonstrate how biased values of
$M/L$ ratios affect the results in the constrained version of the
technique. Contours correspond to the 1-, 2- and 3-$\sigma$ levels. \label{figMC}}
\end{figure*}

In Fig.~\ref{figMC} we present the results of the full spectral fitting of
1800 mock spectra (300 noise realisations for the ages $t$=8, 10 and 12~Gyr;
$Z=-0.3$~dex; $\alpha_1=$1.3 and 2.2, i.e.  close to Kroupa and Salpeter
IMFs) for the spectral resolution $R=7000$ and the signal-to-noise ratio
S/N=100 in the wavelength range 3900$\dots$6800~\AA.  The results are
presented in a way showing the degeneracies between derived stellar
population parameters and the $\alpha_1$ IMF slopes.  The results of the
simulations made using metal-poor stellar populations ([Fe/H]=$-1.0$~dex)
look very similar except the $\sim$30~per~cent higher uncertainties.

Here we see that even the first version of the technique is able to recover
the $\alpha_1$ values with the uncertainty of about 0.1--0.2.  When we
impose the $M/L$ ratio, the precision significantly improves and the
uncertainties of $\alpha_1$ become better than 0.1.  We see the strong
age--$\alpha_1$ degeneracy because the higher mass fraction of low-mass
stars in older populations may compensate the lower derived values of
$\alpha_1$.  The accuracy of about 0.06 well compares to the best available
IMF determination for Galactic open clusters Pleiades
($\Delta\alpha_1$=0.15) and M35 ($\Delta\alpha_1$=0.12) presented in
\citet{Kroupa02}.

In order to test the sensitivity of the second version of our technique to
the biased input $M/L$ values, we performed the simulations increasing and
decreasing the $M/L$ expected from the stellar population models by
10~per~cent.  The results are shown in Fig.~\ref{figMC} by green
($0.9 \cdot M/L$) and cyan ($1.1 \cdot M/L$) contours.  It demonstrates how
$M/L$ systematic errors translate into $\alpha_1$ biases.  One has to keep
in mind that usually the dynamical mass of a real stellar systems is known
with the precision worse than 10~per~cent.

\subsection{The analysis of the distribution of the IMF sensitive information}

\begin{figure*}
\includegraphics[width=\hsize]{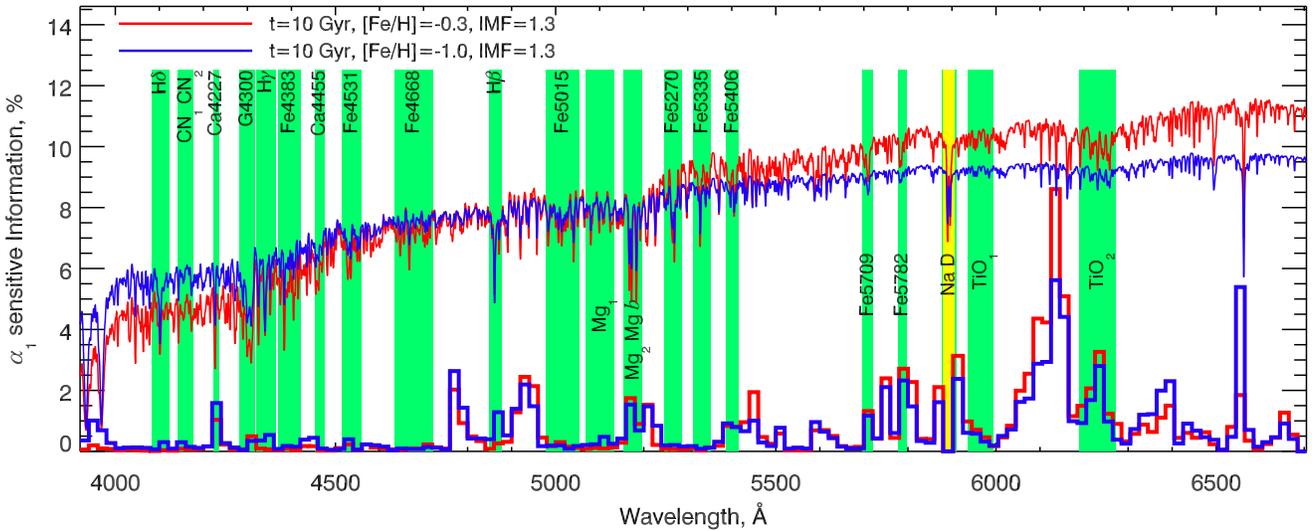}
\caption{The distribution of $\alpha_1$-sensitive information along the wavelength
in 20~\AA-wide bins in the high-resolution (R=7000) spectra of metal-rich
(red, [Fe/H]=-0.3~dex) and metal-poor (blue, [Fe/H]=-1.0~dex) stellar
populations.  The distributions are normalised to the unity (100~per~cent). 
The stellar population spectra are shown above. The Na{\sc i} D region
shown in yellow is excluded from consideration (see the text).
Green areas denote the Lick index definitions. We clearly see that apart 
from a few exceptions, the IMF-sensitive information is not associated with 
Lick indices (see the text). \label{figIMFinfo}}
\end{figure*}

For the practical usage of our technique and future impovements
of the observational strategy to study the IMF in real stellar systems, one
needs to know which spectral features have the highest sensitivity to the
IMF shape.  In order to identify them, we analysed the distribution of the
$\alpha_1$-sensitive information in 20\AA-wide bins in the optical spectral
domain for metal-rich and metal-poor stellar population in a similar fashion
to that proposed in \citet{Chilingarian09} for the analysis of age and
metallicity sensitivity. For our test (see Fig.~\ref{figIMFinfo} we chose
two high resolution ($R=10000$) {\sc pegase.hr} SSP models in the wavelength
range 3920$\dots$6700~\AA\  with the age $t=10$~Gyr and $\alpha_1$=1.3 having
metallicities [Fe/H]=$-1.0$ and $-0.3$~dex.  Then, we convolved them with
Gaussian kernels corresponding to the internal velocity dispersions of
10~km~s$^{-1}$ and thus obtained two model spectra or ``reference SSPs''. 
Then for every of them we varied the $\alpha_1$ value by 0.1.  Later, we
fitted these models against their ``reference SSPs'' using the pPXF
procedure \citep{CE04} with the 10th order multiplicative polynomial
continuum.  The total amount of ``information'' equals to the sum of partial
derivatives of ${{\chi}^2}$ as a function of $\alpha_1$ over all pixels
normalised to 100~per~cent.  The Na{\sc i}~D line region shown in yellow in
Fig.~\ref{figIMFinfo} is excluded from consideration because it is a feature
in stellar spectra heavily affected by the interstellar absorption line and
therefore cannot be reliably modelled in the evolutionary synthesis.

We see that the distributions are very similar for metal-rich and metal-poor
populations (red and blue histograms respectively). An exception is
the H$\alpha$ line (6563 \AA) which contains considerable amount of the information for 
metal-poor populations and is less significant at higher metallicities. Most of the
IMF-sensitive spectral regions are those containing numerous but relatively
faint absorption lines sensitive to the surface gravity in the atmospheres
of late type (GKM) stars.  These lines are concentrated on both sides of
H$\beta$ (4760--4800\AA\ and 4900--4960\AA) and between 5400\AA\ and
6400\AA\ with the most sensitive part around 6100\AA\ associated with Ca{\sc
i}, V{\sc i}, Co{\sc i}, Ni{\sc i}, and Ti{\sc i} absorption lines.  Worth
mentioning that when varying $\log g$, the Ca{\sc i} lines change in the
opposite direction compared to all other elements from the list above. 
Details will be given in a separate paper.

\subsection{Effects of the high-mass IMF shape on the $\alpha_1$
determination}

In this study we use the Kroupa canonical IMF \citep{Kroupa02} by
varying the low-mass end slope $\alpha_1$ and leaving the high-mass slope
fixed ($\alpha_2 = 2.3$).  Recently, \citet{MKDP12} have demonstrated that
stellar systems which underwent high star formation volume densities at a
parsec scale require the introduction of $\alpha_3$ for massive stars ($>1
M_{\odot}$).  Using Monte-Carlo simulations we have checked how changes of
$\alpha_3$ influence the fitting results for a ``standard'' bi-modal
canonical IMF.  We generated 4500 spectra (300 noise realisations for the
ages $t$=4, 6, 8, 10 and 12~Gyr; $\alpha_3$=1.7, 2.0 and 2.6,
$\alpha_1$=1.3, $[$Fe/H$]=-1.0$~dex).  Then we fitted them using the first
version of our technique with a model grid having $\alpha_3=2.3$.  The
results of this fitting are presented in Fig.~\ref{alpha_2,3} as data points
connected by solid lines.

Here we clearly see that the $\alpha_1$ determination remains unaffected by
$\alpha_3$ values for stellar systems older than 8~Gyr.  It is trivially
explained by the fact that the lifetime of massive stars ($>1 M_{\odot}$) is
shorter than 8~Gyr therefore they do not contribute to the integrated light of a
stellar system.  The conclusion from this experiment is that any stellar
system of the age of 8~Gyr or older can be analysed with our standard set of
the SSP models with the fixed value of $\alpha_3=2.3$ when using the first
version of our IMF determination technique.

We performed similar analysis in order to test how the best-fitting
$\alpha_1$ values are affected by the variations of $\alpha_2$.  We
generated 3000 mock spectra (300 noise realisations for the ages $t$=4, 6, 8,
10 and 12~Gyr; $\alpha_2$=2.0 and 2.6, $\alpha_1$=1.3, metallicity
$Z=-1.0$~dex).  Then we fitted them in a similar fashion to the
$\alpha_3$ test.  These results are also shown in Fig.~\ref{alpha_2,3} as
data points connected by dash dot lines.

We see that $\alpha_2$ variations introduce significant biases into
the best-fitting $\alpha_1$ values.  The bias is larger for younger
populations ($\alpha_1$=2.0 instead of 1.3 for 4 Gyr) than for older
populations ($\alpha_1$=1.7 instead of 1.3 for 12 Gyr).  This fact is also
easy to explain by the theory of stellar evolution: for older stellar
populations all massive stars have already evolved into the remnants and the
contribution of stars with masses $M>0.5 M_{\odot}$ decreases as the age
increases.  The lower $\alpha_2$ values correspond to the excess of these
stars compared to the canonical IMF shape and when we fit such a spectrum
with the models computed for $\alpha_2=2.3$, it requires shallower
low-mass end slope $\alpha_1$ to partially compensate the lack of more
massive stars which were present in the model because of lower $\alpha_2$.

However, we stress that according to recent studies there is no evidence for
$\alpha_2$ to vary: ``The IMF of massive stars is well described by a
Salpeter/Massey slope, $\alpha_2$ = 2.3, independent of environment as
deduced from resolved stellar populations in the Local Group of galaxies. 
Unresolved multiple stars do not significantly affect the massive-star
power-law index'' \citep{PK2011}.

\begin{figure}
\includegraphics[width=\hsize]{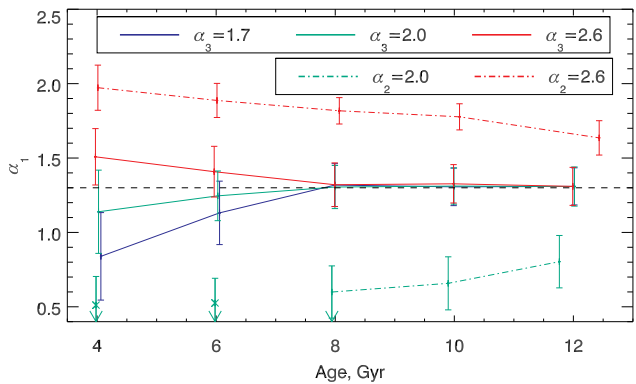}
 \caption{The $\alpha_1$ determination biases introduced by the
variations of $\alpha_2$ and $\alpha_3$.  For every set of parameters, error
bars show the standard deviation of the best-fitting $\alpha_1$ calculated
from 300 noise realisations in every point.  Solid and dash dot lines are
for the fitting results when varying $\alpha_3$ and $\alpha_2$ respectively. 
Arrows indicate that the lower limit of the model grid ($\alpha_1=0.4$) was
reached for some mock spectra.\label{alpha_2,3}}
\end{figure}

\subsection{Full spectral fitting of real data}

We applied our technique to observed spectra having different
signal-to-noise ratios, spectral resolutions and wavelength coverages.  We
used three publicly available datasets for compact stellar systems: (1)
intermediate resolution (R=1300) Galactic GC spectra from
\citet{SRCM05} obtained with the R-C spectrograph at the 4~m Blanco
telescope at CTIO in the wavelength range 3900$<\lambda<$6500~\AA; (2)
high-resolution (R=17000) intermediate signal-to-noise spectra of UCDs in the
Fornax cluster obtained with FLAMES/Giraffe at ESO VLT in the HR9 setup
covering a narrow wavelength interval, 5120$<\lambda<$5450\AA~
\citep{Mieske+08,CMHI11} downgraded to R=10000, and (3) intermediate
resolution (R=1500--2000) high signal-to-noise ratio spectra
(3900$<\lambda<$6800~\AA) of Fornax and Virgo cluster UCDs obtained with
Gemini GMOS-N/GMOS-S \citep{Francis+12}.

The dynamical $M/L$ measurements for some of \citet{SRCM05} spectra
were available from \citet{DMM97} and \citet{Lane+10}.  However, none of the
clusters had $t_{\mbox{rel}} > 4$~Gyr, and those spectra were integrated and
extracted only in the cluster core region.  In addition, classical long-slit
spectra of GCs used for the dynamical analysis and consequently our MF
analysis suffer from systematic uncertainties caused by the limited number
of individual stars falling into the slit (see discussion in
\citealp{DMM97}).  Another problem is that the nuclear regions will have
higher weights over peripheral parts because the cluster light fraction
falling into the slit decreases as $R^{-1}$.  Therefore, we could not obtain
any reliable estimates of $\alpha_1$ values in these objects but only use them
to evaluate statistical errors of the techniques for spectra of such a type. 
The statistical uncertainties of the first version of our technique turned
to be about $\Delta\alpha_1 \approx$0.3--0.5.  Fixing the M/L values reduced them
down to 0.15.

Spectra of extragalactic clusters and UCDs have a number of advantages over
Galactic GC spectra although they are more difficult to obtain because the
targets are much fainter. The main advantage is that these spectra will be
very close to the integrated spectra because at the distance of Virgo and
Fornax CSSs usually stay unresolved from the ground. Therefore, there
will be no systematic errors connected to the low number statistics of stars
on the slit, and also there will be no $R^{-1}$ weighting of peripheral
parts. Another advantage is that in rich environments such as Fornax and
Virgo, the number of GCs is so high that they populate very well even the
brighter end of the luminosity function, therefore it becomes easy to find
objects with $t_{\rm rel} \geq t_{\rm Hub}$ which did not undergo
the dynamical evolution and therefore are suitable for the determination of
IMF and not only the PDMF.

Our experiments with the high-resolution FLAMES/Giraffe spectra of Fornax
cluster GC/UCDs demonstrate that those data can be used for the PDMF
determination \emph{only} if the dynamical $M/L$ values are known from the
analysis of their internal dynamics and structural properties.  Then the
statistical uncertainties of the $M/L$ constrained fitting stay in the range
of 0.05$\dots$0.1 depending on the S/N ratio and the overall errors are
systematic and fully imposed by the $M/L$ uncertainties.  However, very
short wavelength range not including any strong IMF sensitive features makes
the $\alpha_1$ determination impossible for the first unconstrained version
of our technique bringing the statistical error to $\Delta\alpha_1=$0.8 even
for the brightest object, UCD~3 (F~19 in \citealp{CMHI11}).

\begin{figure*}
\includegraphics[width=\hsize]{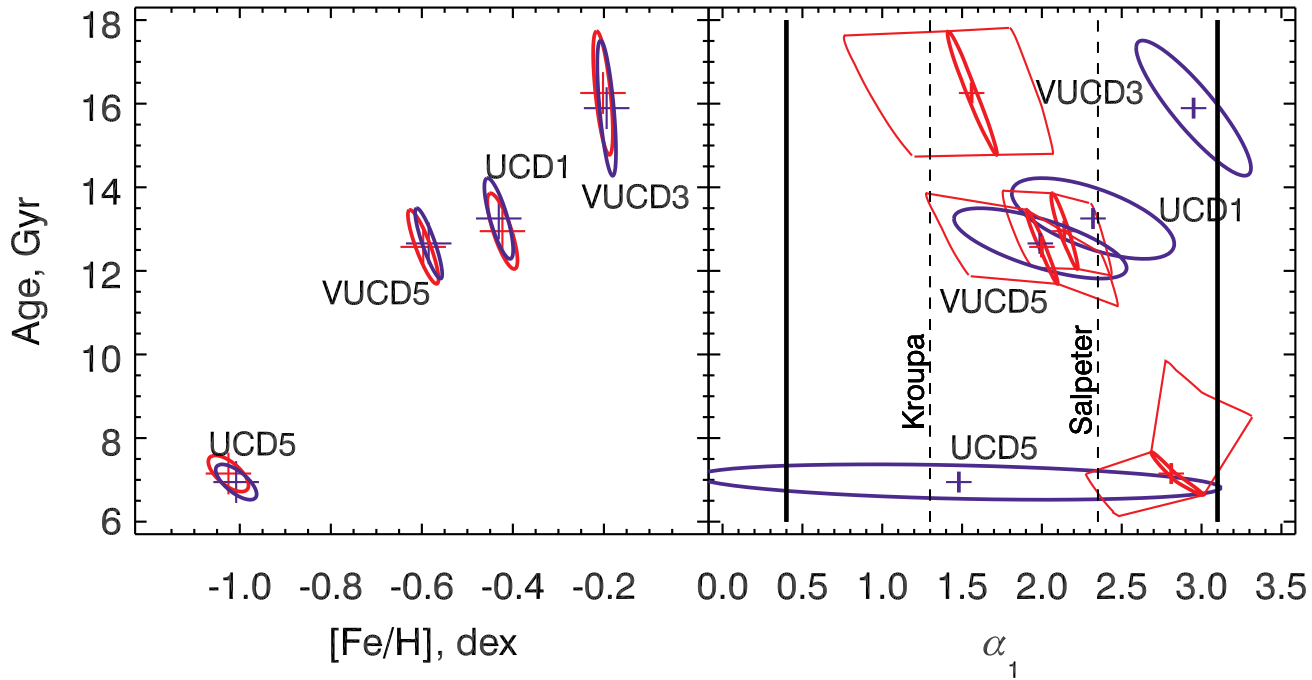}

\caption{Determination of the low-mass IMF slope using both versions of our
technique for four UCDs: UCD~1 and UCD~5 from the Fornax cluster; VUCD~3 and
VUCD~5 from the Virgo cluster.  Blue and red crosses and ellipses denote the
best fitting values and their 1-$\sigma$ statistical uncertainties computed 
from the covariance matrix for the unconstrained and $M/L$ constrained
versions of our technique correspondingly. Red boxes demonstrate the range
of systematic errors for the constrained fitting corresponding to 1-$\sigma$
uncertainties of the adopted $M/L$ ratios. Black lines on the lower panel
show the limits of our grid of SSP models.\label{figGMOS}}
\end{figure*}

Analysis of high S/N integrated spectra for Virgo/Fornax UCDs obtained with
Gemini GMOS-N and GMOS-S spectrographs in a broad wavelength range
demonstrates a reasonably good agreement between the two versions of our
technique (see Fig~\ref{figGMOS}).  We used published dynamical $M/L$
ratios for two Virgo cluster UCDs (VUCD~3 and VUCD~5,
\citealp{Evstigneeva+07}) and two Fornax cluster UCDs (UCD~1 and UCD~5,
\citealp{CCB08}).  The adopted $(M/L)_{V,\mbox{dyn}}$ values in the Solar 
units are: $5.4 \pm 0.9$, $4.1 \pm 0.8$, $5.0 \pm 0.6$, and $3.4 \pm 0.9$ 
for VUCD~3, VUCD~5, UCD~1, and UCD~5 correspondingly.
For UCD~1 and VUCD~5 both, fully unconstrained fitting
and the $M/L$-imposed version of the technique agree well within the
statistical uncertainties of the unconstrained fitting.  For UCD~5 having
low metallicity and therefore much weaker absorption lines, the
unconstrained fitting results in very large uncertainties while the adopted
$(M/L)_V$ value puts it close to the edge of the model grid
on the MF slope (2.8).  For VUCD~3 the situation is the opposite, the
unconstrained fitting yields $\alpha_1 \approx 2.9$, however one has to keep
in mind that this object has extremely high value of the [$\alpha$/Fe]
chemical abundance ratio \citep{Francis+12} resulting in the significant
template mismatch, therefore the unconstrained measurements may get biased. 
At present, we do not have models required to study the biases which may be
introduced to the MF measurements from integrated light spectra by
significantly non-solar abundance ratios.

\section{Discussion and Summary}

We presented a novel technique for the determination of the low-mass end
slope of the PDMF of unresolved stellar populations using pixel fitting of
spectra integrated along the line of sight.  In stellar systems with long
dynamical relaxation timescales such as massive GCs, UCDs and galaxies,
where one can neglect the effects of the dynamical evolution, the PDMF is
identical to the IMF at low masses.  Our technique is available in two
versions having different applications: (a) fully unconstrained
determination of age, metallicity and the low-end PDMF slope $\alpha_1$
which can be used to analyse high signal-to-noise spectra of CSSs and normal
galaxies given that the wavelength coverage includes IMF sensitive features;
(b) constrained determination of the same parameters by imposing the
external value of the $M/L$ ratio which can be, for example obtained from
the dynamical modelling and therefore assuming the zero dark matter content
in a stellar system being studied.

We used Monte-Carlo simulations with mock spectra to test both approaches
and conclude that: (1) age, metallicity, and IMF can be precisely determined
in the first unconstrained version of the code for high signal-to-noise
datasets (e.g.  S/N=100, R=7000 give the uncertainty of $\alpha_1$ of about
0.1--0.2); (2) adding the $M/L$ information significantly improves the
precision and reduces the degeneracies, however systematic errors in the
imposed $M/L$ ratio will translate into offsets in $\alpha_1$.  

We performed the analysis of the IMF-sensitive information distribution
across the optical spectral domain.  One can clearly see in
Fig.~\ref{figIMFinfo} that apart from a few exceptions, IMF sensitive
information is not associated with prominent absorption line features (and
line strength indices corresponding to them, e.g.  Lick indices, see
\citealp{Worthey94}) where a significant fraction of age- and
metallicity-sensitive information is contained.  The exceptions are: the
Ca{\sc i} feature ($\lambda = 4227$~\AA) and the Mg$b$ triplet ($\lambda
\approx 5172$~\AA).  However, these two indices are very sensitive to the
metallicity and the [$\alpha$/Fe] abundance ratios and therefore can hardly
be used to constrain the IMF.  Another IMF sensitive Lick index is
Fe5782 which is known to be difficult to measure and calibrate.  The only
strong IMF feature left is the TiO$_2$ index including about 7~per~cent of
the total IMF-sensitive information.  Hence, ``standard'' absorption line
strength indicators cannot be used to determine $\alpha_1$.

According to the recent studies \citep{MKDP12}, the stellar IMF at high
masses ($>1 M_{\odot}$) appears to depend on the star-formation rate density
on a parsec scale and even becomes top-heavy when it surpasses
0.1$M_{\odot}$ / (${{\rm yr}\cdot {\rm pc}}^{3}$).  It may become
increasingly bottom-heavy at high metallicity.  The increased number of
compact stellar remnants such as neutron stars and black holes for a
top-heavy IMF will elevate the $M/L$ values \citep{DKB09} and mimic the
dark matter in CSSs.  Conversely, for a top-light IMF the $M/L$ will be lower
then for the canonical Kroupa IMF. This will become a caveat for the
second ($M/L$ constrained) version of our technique because the $M/L$ ratios
computed from the stellar population models based on the Kroupa canonical
IMF will be incorrect. We quantified this effect by comparing
the {\sc pegase.2} predictions of stellar $M/L$ ratios for stellar populations
having different $\alpha_1$, $\alpha_3$, and ages.  As expected, the
$\alpha_3$-induced $M/L$ variations are the most important for bottom-light
populations ($\alpha_1 < 1$) because of the higher relative mass fraction of
stellar remnants.  However, the overall effect of the $\alpha_3$ variations
on $M/L$ ratios is lower than that of $\alpha_1$ for the same amount of the
IMF slope change.  For instance, going from $\alpha_3=2.3$ to $\alpha_3=1.7$
changes the $V$ band $M/L$ ratio by 11 (for $\alpha_1=3.1$) to 26~per~cent
(for $\alpha_1=0.4$) for ages exceeding 10~Gyr.  For 8~Gyr the effect is
even 2~per~cent smaller.  The top-light IMF ($\alpha_3=2.6$) lowers the
$M/L$ by as little as 1 to 6~per~cent for ages above 8~Gyr.

At the same time, the unconstrained version of our fitting technique will
provide correct estimates of the low-mass IMF slope if applied to older
stellar populations {($t>8$~Gyr)} where massive stars affecting the variable
high-end IMF slope have already evolved into stellar remnants (see section
2.3).  It also can be applied to the data for galaxies because the assumed
dark matter content does not have to be zero.  The unconstrained
version can be applied to check whether the IMF is bottom heavy in massive
elliptical galaxies where low-mass stars might have efficiently formed in
cooling flows \citep{KG1994}.  However, we have to keep in mind that the
difference between SSPs and real star formation histories of galaxies as
well as significantly non-solar element abundance ratios may introduce some
biases.

We applied our technique to observed spectra of CSSs and demonstrated that
even for low resolution datasets ($R=1500$) with high S/N-ratios the IMF
slope can be precisely determined even using the unconstrained fitting of
$\alpha_1, t, [$Fe/H$]$.  The $M/L$ constrained approach provides even
higher precision and agrees well with the unconstrained version for objects
without peculiarities in stellar populations and chemical abundances.

High spectral resolution will improve the precision only if IMF-sensitive
spectral features are covered.  Extragalactic GC systems where structural
parameters of clusters are available from the Hubble Space Telescope data
will have advantage over the Galactic GCs because the integrated velocity
dispersion measurements of such systems are much easier to obtain for large
samples of targets using multi-object spectroscopy.
	
We conclude that by applying our technique to high-quality optical
observations of CSSs we are able to reach better precision of the IMF
determination than that made with direct star counts in nearby open clusters
and check the IMF universality hypothesis.

\section*{Acknowledgments}

The work was supported by the Russian Federation President's grant
MD-3288.2012.2, Russian Foundation for Basic Research (project no. 
12-02-31452) and M.~V.~Lomonosov Moscow State University Program 
of Development. We thank our anonymous referee for valuable comments. 
IYK is grateful to Dmitry Zimin's non-profit Dynasty Foundation.

\bibliographystyle{mn2e}
\bibliography{imf_method}

\label{lastpage}

\end{document}